\RequirePackage{fix-cm}
\documentclass[smallextended]{svjour3}       
\usepackage{graphicx}
 \usepackage{epstopdf}

\usepackage{color}

\begin{document}

\title{Testing for rational speculative  bubbles in the Brazilian residential real-estate market
}


\author{Marcelo M. de Oliveira       \and
        Alexandre C. L. Almeida 
}


\institute{
              Authors afilliation: Universidade Federal
de S\~ao Jo\~ao del Rei, Campus Alto Paropeba, \\
36420-000, Ouro Branco, Minas Gerais - Brazil\\
\\
              \email{mmdeoliveira@ufsj.edu.br} (M.M. de Oliveira)       \\
 \email{celestino@ufsj.edu.br} (A.C.L.  Almeida) \\
}

\date{Received: date / Accepted: date}

\maketitle


\vspace{2cm}

\begin{abstract}
Speculative bubbles have been occurring periodically in local or global real estate markets and are considered a potential cause of economic crises. In this context, the detection of explosive behaviors in the financial market and the implementation of early warning diagnosis tests are of critical importance. The recent increase in Brazilian housing prices has risen concerns that the Brazilian economy may have a speculative housing bubble. In the present paper, we employ a recently proposed recursive unit root test in order to identify possible speculative bubbles in data from the Brazilian residential real-estate market. The empirical results show evidence for speculative price bubbles both in Rio de Janeiro and S\~ao Paulo, the two main Brazilian cities.
\vspace{0.5cm}
\end{abstract}

\newpage
\section{Introduction}
\label{intro}
The efficient market hypothesis (EMH) assumes that all information is reflected
instantaneously in asset prices \cite{fama}. Hence, market prices should always be consistent with the
``fundamentals". On the other hand, sudden dramatic price changes over short periods of time, in which prices diverge from fundamentals, have been observed in several markets around the world \cite{Roehner02}. Famous examples include the Dutch ``Tulipmania", the 1929 Crash in the United States and the recent American subprime crises \cite{garber,kindleberger}.
Such behavior has raised many questions regarding market efficiency and has stimulated research on rational speculative bubbles \cite{evans,cajueiro}.

Rational speculative bubbles occur when there is an excessive public expectation of future price increasing, which produce rapid increases in valuations of an asset \cite{abreu}. For example, during a real-estate bubble, investors (homebuyers) stay in the market despite the deviations of prices from fundamentals because they expect  significant further price increases. Additionally, when the bubble is emerging, people think that real estate prices are very unlikely to fall, letting to a little perceived risk associated with investing in the real estate market. But eventually the prices may reach unsustainable levels, and crash: the bubble bursts \cite{abreu,sorne03b,sorne03}. 

Bubbles, rational or not, have been occurring periodically in local or global real estate markets and are considered of critical importance and a fundamental cause of financial crises and ensuing economic crises \cite{Roehner02,roehner06,richmond,mayer}. Hence, the study of real-estate bubbles is an important contribution to the literature on future economic development. Moreover, a bubble is hazardous for financial and macro stability, since it can amplify a credit boom by inflating collateral values and causing misallocation of economic resources. In order to minimize the
negative macroeconomic effects of a bubble, one needs to detect it early – as soon as prices begin to rise.

The usual definition of a bubble is a deviation of the asset price from its fundamental value. However it can be very difficult to evaluate what is the ``fundamental" price. Alternatively, constantly diminishing dividend-price ratio can serve as a
reference for identifying a rational bubble in a stock market. If price expectations are rising, but higher dividends fail to
materialize, the price rise is probably not based on fundamentals. A large number of studies have been proposed to empirically detect asset bubbles, mostly focusing in stock markets. Some of these methods have been applied to the study of real estate bubbles. Econometrics tests, like unit root tests or cointegration tests provide direct tests for the no-bubble hypothesis. If property price to  income (or rent) ratio is stationary or the property price is cointegrated with the fundamental price, the no-bubble hypothesis cannot be rejected. 

The first econometric tests for rational bubble detection were volatility tests such as Schiller's \cite{variance} variance-bound and West's two-step test \cite{west}. The underlying idea is to compare the volatility between the assumed
fundamental asset price and the actual asset price. An asset bubble is detected indirectly if the volatility of the actual asset price is significantly larger
than that of the assumed fundamental asset price. However, Marsh and Merton \cite{merton} pointed evidence that variance bounds test fails when dividends and stock prices are non-stationary. 

Campbell and Shiller \cite{camp-shi} and Diba and Grossman \cite{diba} introduced the most commonly used methods for detecting asset price bubbles in the literature, namely the right-tailed unit root test and the cointegration test. The aim of a cointegration test is to  examine whether or not variables are trending together. If two series are cointegrated, that means the movements of these two variables are highly correlated. In other words, there is an equilibrium relationship between these two series. Related tests using data sets that combine time series and
cross-sections (Panel-based tests) have been applied recently in order to investigate the existence of real estate bubbles in Taiwan \cite{chun}, China \cite{hui}, as well USA and Europe \cite{taipalus}. These methods, however, suffer from a serious limitation, first
pointed-out by Evans \cite{evans}, who showed that cointegration tests of asset prices and dividends are not able to detect explosive
bubbles when the sample data includes periodically collapsing bubbles.

A novel approach to identifying and dating bubbles in real time has recently been introduced by Phillips and Yu \cite{phillips11a}.  Considering that the explosive property of bubbles is very different from random walk behavior, they developed a recursive econometric methodology interpreting mildly explosive unit roots as a hint for bubbles. The method provides a supremum Dickey-Fuller (DF) test \cite{ADF} that overcomes the problem identified with unit root and cointegration tests. The supremum DF test improves power significantly with respect to the conventional unit root and cointegration tests, and
has the advantage of allowing estimation of the origination date and final date of a bubble. The idea is to detect speculative bubbles as they emerge, not just after their burst. This methodology was generalized in \cite{phillips12} for detecting multiple bubbles. Yiu et al \cite{yu} applied the latter for detecting
bubbles in the Hong Kong residential property market, Chen and Funke \cite{funke} for the Chinese housing market and Gonz\'alez {\it et al.} \cite{borradores} for the Colombian housing market. A related approach was employed by Kideval \cite{kideval} for detecting rational bubbles in the US housing market.

It is noteworthy to mention a few tests for bubbles not based on econometric approaches. Sornette {\it et al.}\cite{sorne97} intend to predict the end of the bubbles assuming that a crash follows after rapid growth of economic indicators faster than an exponential function. Zhou and Sornette \cite{sorne06,sorne08} examined whether the price bubble burst in the US, and predicted that the turning point of the bubble would occur around mid-2006. Watanabe {\it et al.} \cite{watanabe,hui2} propose to identify bubbles and crashes by exponential behaviors detected in the systematic data analysis. More recently, Ohnishi {\it et al.} \cite{plaw} observed that the land
price distribution in Tokyo had a power law tail during the bubble period in the late
1980s, while it was very close to a lognormal before and after the bubble period.  This led them to argue that a characteristic of real
estate bubbles is not the rapid price hike itself but a rise in cross-sectional dispersion of prices.

Over the past few decades, the real estate market has been a target of government fiscal and monetary policies aimed at achieving balanced economic growth, low unemployment and low inflation. When the housing market or the overall economy is on a downturn, the governments tend to encourage banks to adopt a lenient mortgage policy, increasing the available credit as well to implement policies favorable to construction companies. Another usual government policy consists on subsiding disadvantaged groups through preferred loan rates. As a consequence, the macroeconomic market is filled with speculative capital demand and supply, resulting in inefficiency in the operation of various markets. This situation creates the conditions to the onset of a 
bubble \cite{chun}. 

Such situation is exemplified by the Brazilian real estate policies. At the end of 2007, the world was affected by the subprime mortgage crisis in the United States.  In response, Brazilian government established various market stimulus policies \cite{pac}, which includes: (i) Approval of a new law of fiduciary alienation that minimizes the risk for the Brazilian banks in case buyers default on their loans. This make the banks more willing to lend to potentially riskier buyers. (ii) Brazil Growth Acceleration Program (referred to as PAC) which included loans provided by government-owned banks, as well long-term credit for infrastructure, sanitation improvements, and more under a new investment fund that began with 657,4 billion Brazilian Reals (BRL)\footnote{During the period 2007-2012 the value of US dollar (USD) to BRL rate was in the range 1.55 to 2.46, with a mean of about 2.00.Currently 1.00 BR $\sim$ 2.30 USD.} in the period 2007-2010 and more 955,1 billion BRL for the period 2011-2014 . (iii) Programs administered by government-owned banks with the aim of development of 1,000,000 houses and apartments with subside for poor families, resulting in a credit of  96 billion BRL in 2011, 18 times the amount of credit available in 2003. 

In fact, real estate prices in Brazil have raised rapidly in the last few years, and some analysts have been arguing the possibility that a bubble has been inflated and could potentially burst \cite{manarin,financial,gamble,sashida,gaulard}. In contrast, others argue that the pricing growth is sustainable and based on fundamentals, since Brazil was one of the fastest-growing major economies in the world in recent years with an average annual Gross Domestic Product (GDP) growth rate of over 5 percent, which made Brazilian economy the world's seventh largest by nominal GDP by the end of 2012. As property is a sizable component of household and corporate balance sheets, a sudden collapse in property prices may have negative spillover effects on the overall macroeconomic situation and may pose macroeconomic and financial stability risks.

Therefore, the aim of this paper is to investigate the bubble-like
behavior of the recent real estate market prices in Brazil.
Identifying speculative bubbles is not an easy task even in mature markets with long time series. Since in Brazil the time series for house prices are short, the recent developed test by Phillips {\it et al.} \cite{phillips12}, aimed at identifying explosive bubbles in real time, provides an adequate tool for such analysis \cite{homm}. 


\section{Methodology}

Identifying speculative bubbles is a hard task even in data sets with long time series.
Recently, based on previous works of Shiller \cite{shiller2000}, Mikhed and Zemcik \cite{mikhed09} regarded a house as an investment asset and used a standard present-value formula to derive implications for the relationship between house prices and the cash flow associated with owning a property (rent).

The fundamental price is derived from the standard no arbitrage condition:

\begin{equation}
\label{eq:Pt}
P_t=\frac{E_t\left[R_{t+1}+P_{t+1}\right]}{1+r}
\end{equation}
where $P_t$ is the property price index at period $t$, $E_t\left[.\right]$ denotes the {mathematical} expectation conditional on information at time $t$, 
$R_t$ is the rent, and $r$ is a constant risk-free discount rate. Since the formula above
holds for all $t$, the property price index for the time $t+1$ is given by

\begin{equation}
P_{t+1}=\frac{E_t\left[R_{t+2}+P_{t+2}\right]}{1+r}.
\end{equation}

Therefore, the Equation \ref{eq:Pt}, by repeated forward iteration, can be written as
\begin{equation}
\label{eq:PtFF}
P_t=E_t\left[\frac{R_{t+1}}{1+r}+\frac{R_{t+2}}{(1+r)^2}+...+\frac{R_{t+k}}{(1+r)^k}+\frac{P_{t+k}}{(1+r)^k}\right].
\end{equation}

{
Solving the Equation \ref{eq:PtFF} yields the fundamental price:
\begin{equation}
P_t^f=\sum_{j=1}^{\infty}\frac{1}{(1+r)^j}E_t\left[R_{t+j}\right]
\end{equation}
also called price reflecting fundamentals.  This equation means that the fundamental price contain all expected future rents.
In absence of a bubble, one has the no-bubble condition \cite{mikhed09}:
\begin{equation}
\label{eq:noBubble}
\lim_{k\to\infty}{\frac{E_t\left[P_{t+k}\right]}{(1+r)^k}}=0.
\end{equation}
This yields that the unique solution of Equation \ref{eq:Pt} is $P_t=P_{t}^{f}$ under the hypothesis of non existence of a bubble.
}
%

The spread $S_t$ between the house price and rent can be defined \cite{campbell,mikhed09} as
\begin{equation}
S_t\equiv P_t-\frac{1}{r}R_t,
\end{equation}
which, based in the no-bubbles condition (Equation \ref{eq:noBubble}), can be rewritten as

{
\begin{equation}
S_t=\frac{1}{r}E_t \left[ \sum_{j=1}^{\infty}\frac{\Delta R_{t+j+1}}{(1+r)^j}\right]=\frac{1}{r}E_t\left[\Delta P_{t+1}\right]
\end{equation}
}
{where $\Delta R_{t+j+1}=R_{t+j+1}-R_{t+j}$ and $\Delta P_{t+1}=P_{t+1}-P_{t}$.}
Note that the stationarity of $S_t$ implies the 
{series $\{P_t/R_t\}$ is}
 stationary  in the absence of a speculative bubble, since 
$P_t/R_t=1/r$ if $S_t=0$. 

In order to estimate the fundamental value of property prices, we will employ the price-rent ratio, $PR$. The $PR$ ratio is defined as the average cost of ownership divided by the estimated rent that would be paid if renting:
\begin{equation}
PR=\frac{\mbox{house price}}{\mbox{annual rent}}.
\end{equation}
The  PR ratio follows the concept of stocks' price-earnings ratio (PER), which is defined as the ratio of the price of a share to the annual earnings of a company in the current year. The PER ratio contains information about if a given stock is over (or under) valuated \cite{campbell,shiller2000}. Analogously, rents, as well as corporate and personal incomes, are usually connected very close to supply and demand fundamentals. This is the cause one rarely sees an unsustainable ``rent bubble" or an unsustainable ``income bubble". Therefore, a rapid increase of housing prices combined with a flat or slow-increasing renting market can be a signal of the beginning of a bubble \cite{gallin}. 
{As mentioned previously, in the absence of bubble the series $PR_t$ is stationary. The way to ascertain whether or not bubbles exist is
testing the stationarity of the house price-to-rent ratio.}
{This leads us to seek methods to determine whether  series are stationary or not}.


Unit root tests are commonly used to determine whether a time series is stationary by using an autoregressive model. In its simplest form, the Dickey-Fuller test \cite{ADF} estimates the following first order autoregressive AR(1) regression equation:
\begin{equation}
\Delta p_t = \alpha+ (\beta)p_{t-1}+\epsilon_t, \epsilon_t\sim iid(0,\sigma^2).
\end{equation}
where ${p_t}$  is the real price of the asset, $\alpha$ is the drift  and $\beta$ is the coefficient of the model. 
The error term $\epsilon_t$ is an uncorrelated white noise process. The DF test compares the t-statistics of residuals
with DF critical values. The null hypothesis of the test is $H_0 : \beta = 0$
which represents unit root versus the left-tailed alternative hypothesis $H_1 : \beta < 0$ stable root. (If the residuals in first order autoregressive model are still correlated the test can be augmented by $\Delta P_{t-i}$ for higher level autoregressive processes).

In contrast, the supremum augmented DF (SADF) test proposed by {Phillips and Yu  \cite{phillips11a} } is a right-sided test.
{The basic fundamental of this test is using recursive regression techniques to test the unit root.}
 In the context of DF test, the test is based on the follow regression:
\begin{equation}
\label{eq:H0}
\Delta p_t = \alpha+ (\beta-1)p_{t-1}+\epsilon_t, \epsilon_t\sim iid(0,\sigma^2).
\end{equation}

The null hypothesis still is unit root behavior, $H_0:\beta=1$ , but the alternative hypothesis is explosive behavior, $H_1:\beta>1$. 
The right-tailed ADF statistics is computed in multiple recursive regressions for each sub-sample which start with the initial observation but the last point varies. 
Let $r_1$ and $r_2$  be respectively the fractional starting and ending point of each sample. The sample window $r_w=r_2-r_1$ therefore varies from the initial size window $r_0$ to the total sample.

The SADF statistic is based on the supremum value of the ADF statistics obtained,

\begin{equation}
SADF(r_0)= \underbrace{\mbox{sup}}_{r_0 \leq r_2\leq 1} ADF_0^{r_2}.
\end{equation}

The explosiveness of the process is tested  by comparing with the right-tailed critical values of its limit distribution which is given by


\begin{equation}
\underbrace{\mbox{sup}}_{r_0 \leq r_2\leq 1} ADF_0^{r_2} \to  \underbrace{\mbox{sup}}_{r_0 \leq r_2\leq 1}\frac{\int_0^{r_2}WdW}{\int_0^{r_2}W^2dW}
\end{equation}
where W is a standard Wiener process {and $\to$ denotes convergence in distribution}.

Instead of fixing the starting point of the sample, the generalized SADF (GSADF) test 
{extends the sample sequence by changing both the starting 
and the ending point of the sample}, implementing the right-tailed unit root test repeatedly on a forward expanding sample sequence:


\begin{equation}
\displaystyle GSADF(r_0)=\underbrace{\mbox{sup}}_{\stackrel{r_0 \leq r_2\leq 1}{0 \leq r_1\leq r_2-r_0}} ADF^{r_2}_{ r_1}.
\end{equation}

{The GSADF statistics can be defined as the largest ADF statistic over the feasible ranges of $r_1$ and $r_2$.}
It is then used to detect the presence of at least one bubble in the whole sample.
{Phillips {\it et al.}(2012) \cite{phillips12} } demonstrate that the moving sample GSADF diagnostic outperforms the SADF test in detecting explosive behavior in multiple bubble episodes, and works well even in modest sample sizes.

In order to estimate the beginning and collapse dates of every bubble, Phillips {\it et al.} (2012) suggest using backward expanding sample sequences.
Let the fractional ending point fixed at $r_2$ with the starting point $r_1$ moving in the range $0\leq r_1\leq r_2-r_0$. 
{These ADF statistic sequences are denoted by }
\begin{equation}
\left\{BADF_{r_1}^{r_2}\right\}_{0\leq r_1\leq r_2-r_0}. 
\end{equation}
Hence the backward SADF statistic is defined as the sup value of the ADF
statistic sequence:


\begin{equation}
BSADF_{r_2}(r_0)=\underbrace{\mbox{sup}}_{0\leq r_1 \leq r_2-r_0}\left\{ BADF_{r_1}^{r_2}\right\}.
\end{equation}

The beginning (end) date of a bubble corresponds to the {first} date 
whose the 
BSADF statistic
becomes greater (smaller) than the 
{critical values, estimated} by Monte Carlo simulation. {(These critical values are calculated from respective empirical distributions of each statistic, generated under the null hypothesis (Equation \ref{eq:H0} with $\beta=1$) )}

\section{Brazilian residential real estate market}

\begin{figure}[ht]
\begin{center}
\includegraphics[clip,angle=0,scale=0.6]{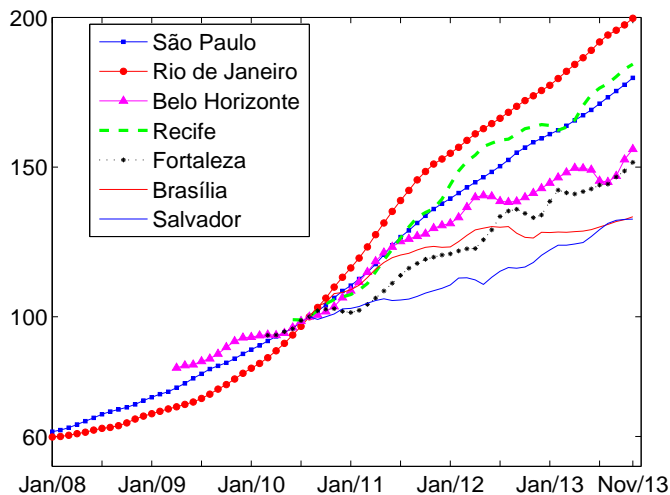}
\end{center}
\caption{Fipe-Zap index for the seven biggest Brazilian cities.(Color
  online).}
\label{fig:precos}
\end{figure}

Prior to the econometric analysis, let us briefly describe the data set used. Until recently, Brazil had no reliable indicators following the price behavior of residential properties. For this purpose, since 2008, Fipe (Brazilian Institute of Economic Research) developed an index, named Fipe-Zap, which use real estate ads as a source of information. The major drawback with this source is a possible gap between the advertised price and the realized price. Nonetheless, if one assumes that both prices have a similar trend, at least in the medium and long run, an index for the advertised prices could be regarded as reliable real estate market information. The index is based on a database of approximate 200
thousands real estate ads per month, and its methodology is detailed in \cite{fipezap}. Figure     \ref{fig:precos}        shows the price index for the seven biggest Brazilian cities (for each index it was set an arbitrary value of 100 on August 2010). Almost all the indices show high increases during the period, revealing that Brazilian properties prices rose rapidly in the last few years. Unfortunately, only the data for the two major cities, S\~ao Paulo (SP) and Rio de Janeiro (RJ), is available since December 2007. The sample data is also larger for these both cities, therefore for a more precise statistical analysis, from now on we will concentrate our analysis on the SP and RJ real-estate market.  

\begin{figure}[ht]
\begin{center}
\includegraphics[clip,angle=0,scale=0.6]{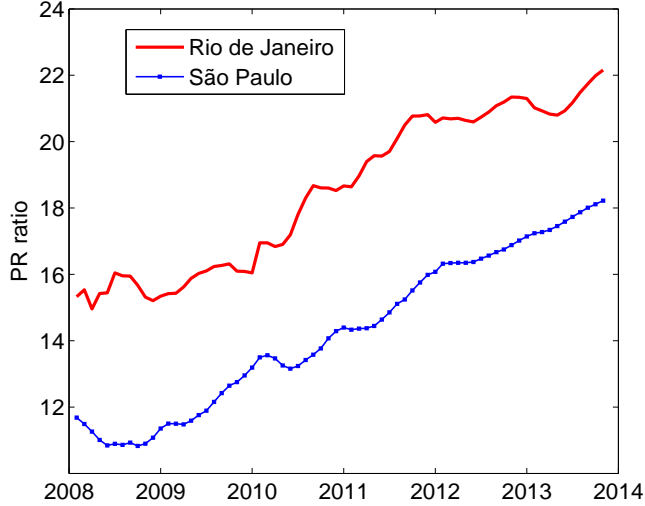}
\end{center}
\caption{Price-rent ratio for S\~ao Paulo and Rio de Janeiro.(Color
  online).}
	\label{fig:PRratio}
\end{figure}

Since Fipe also calculates the monthly rental yield from its rent and sale ads database, it is straightforward to obtain the PR ratio as
the reciprocal of the annualy rental yield. The resulting time-series with the PR ratio for SP and RJ is shown in Fig. \ref{fig:PRratio}. 
 In SP, we observe that the PR ratio grows from around 11 to 18, and in RJ from around 15 to 22. This is equivalent to a increase of approximately  $64\%$ and $47\%$, respectively. It should be noted that a rising PR ratio is only a necessary but not a sufficient 
condition for speculative misalignment from fundamentals. 

\section{Empirical Results}

In order to test whether the movement of house prices in the Brazilian real-estate market reflects deviation from
levels supported by fundamentals we performed more detailed analysis of the data shown in Fig.  \ref{fig:PRratio}. Thus, we applied the recursive SADF and GSADF tests to the PR dataset. In our analysis, we choose the minimal window size $r_w=12$ which ensures that there are enough observations for initial estimation. The finite sample critical values were obtained via Monte Carlo simulations with 10,000 iterations. The resulting SADF and GSADF statistics for the Rio and S\~ao Paulo PR ratio are shown in Table \ref{tab:stat}. The SADF statistics indicate the existence of at least one speculative real estate bubble in Rio de Janeiro during the period with a  $95\%$ confidence interval and in S\~ao Paulo with a $99\%$ confidence interval. Similar conclusions can be drawn from the GSADF test statistics, with bubble in Rio de Janeiro (at $90\%$ level) and S\~ao Paulo (at $99\%$  level).

\begin{center}
\begin{table}
\caption{SADF and GASDF statistics for PR ratio series and respective 90\%, 95\% and 99\% critical values.}
\label{tab:stat}
\begin{tabular}{ccccc}

\hline\hline PR ratio & SADF $t$-statistics  & 90 c.v. & 95 c.v. & 99 c.v.\\
\hline 
Rio de Janeiro& 1.3203 &1.0106& 1.3144 & 1.9812 \\
S\~ao Paulo &  2.6529  &1.0106& 1.3144 & 1.9812\\
\hline 
& GSADF $t$-statistics  & 90 c.v. & 95 c.v. & 99 c.v.\\
\hline
Rio de Janeiro& 1.9180 &1.8312 & 2.1804 & 2.9606 \\
S\~ao Paulo &  4.1353 &1.8312 & 2.1804 & 2.9606\\
\hline \hline

\end{tabular}
\end{table}
\end{center}

\begin{figure}[ht]
\begin{center}
\includegraphics[clip,angle=0,scale=0.6]{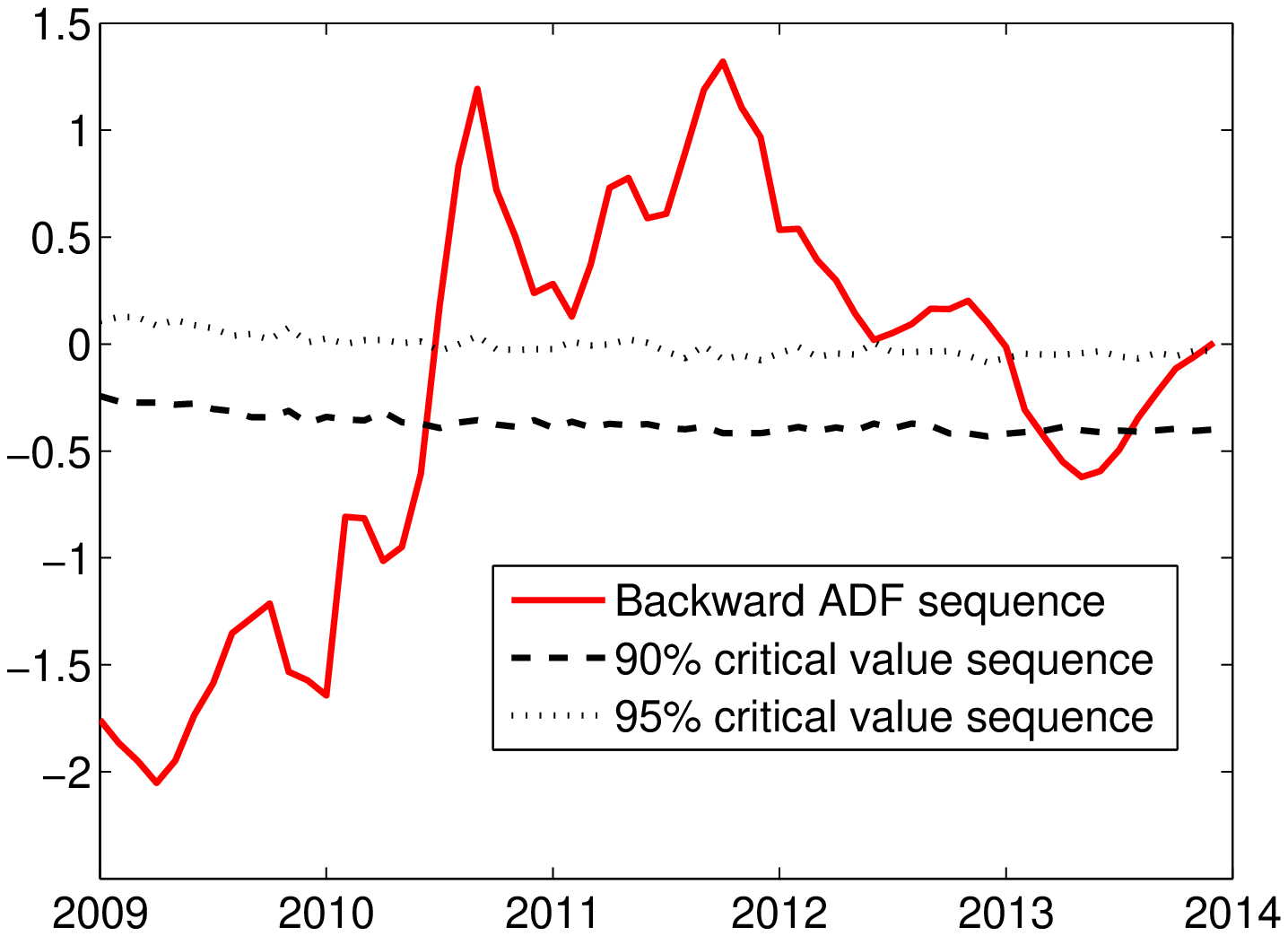}
\end{center}
\caption{Recursive calculation of the SADF test for Rio de Janeiro. The doted lines represents the  95\% and the dashed 90\% critical value sequences (Color  online).}
\label{fig:RJadf}
\end{figure}

\begin{figure}[ht]
\begin{center}
\includegraphics[clip,angle=0,scale=0.6]{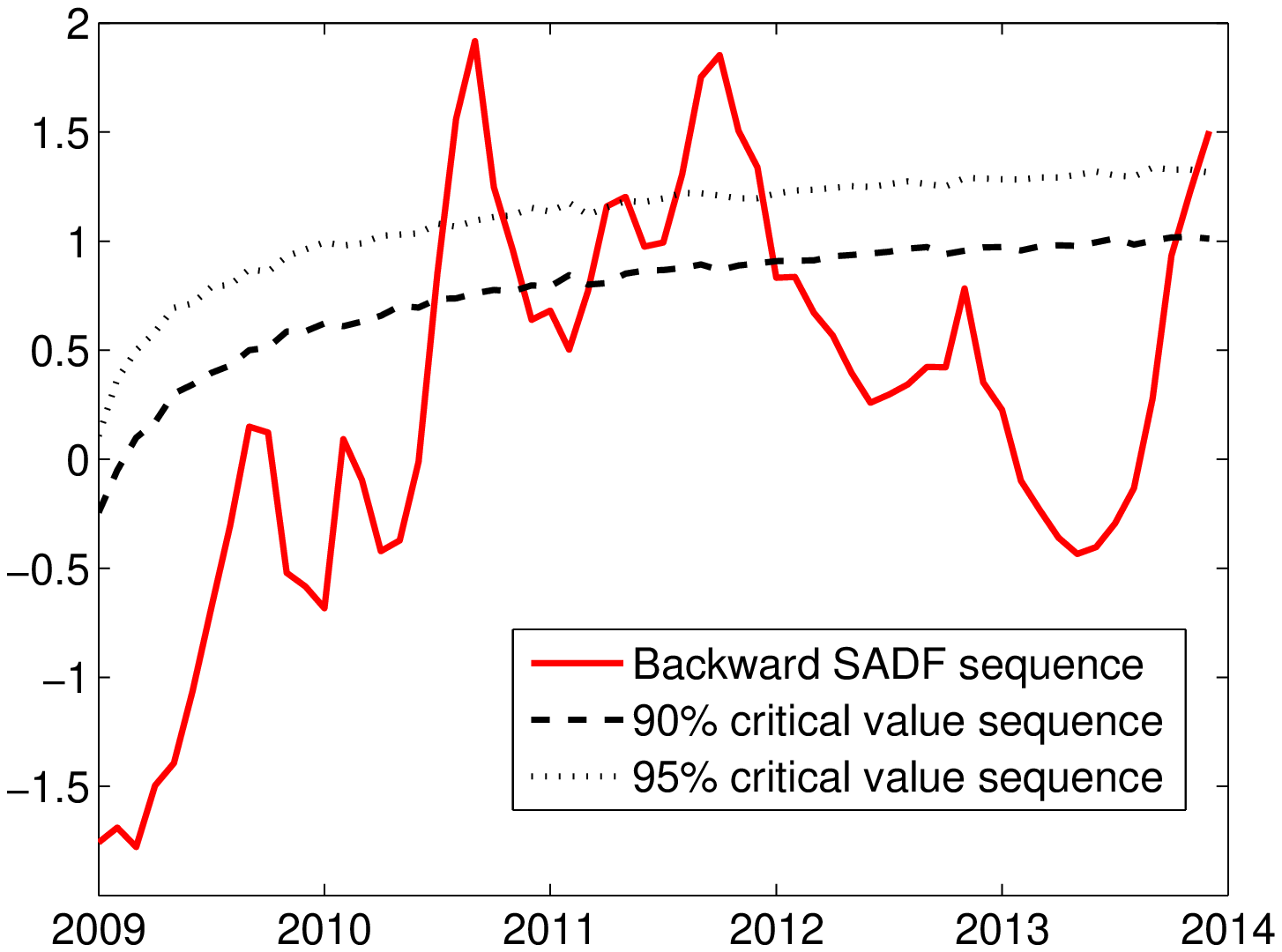}
\end{center}
\caption{Recursive calculation of the GSADF test for Rio de Janeiro. The doted lines represents the  95\% and the dashed 90\% critical value sequences (Color  online).}
\label{fig:RJsadf}	
\end{figure}

\begin{figure}[ht]
\begin{center}
\includegraphics[clip,angle=0,scale=0.6]{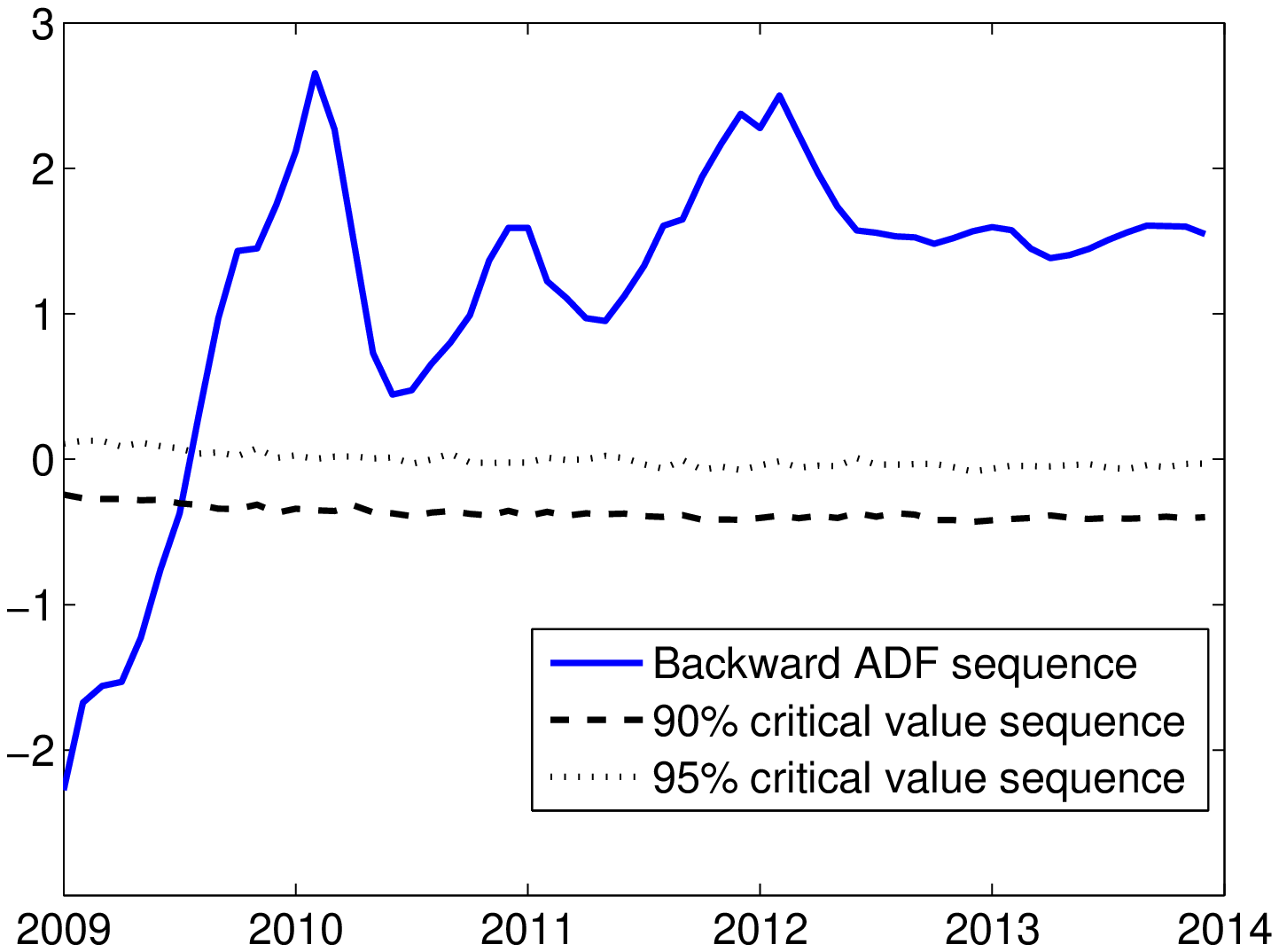}
\end{center}
\caption{Recursive calculation of the SADF test for S\~ao Paulo.The doted lines represents the  95\% and the dashed 90\% critical value sequences (Color
  online).}
	\label{fig:SPadf}
\end{figure}

\begin{figure}[ht]
\begin{center}
\includegraphics[clip,angle=0,scale=0.6]{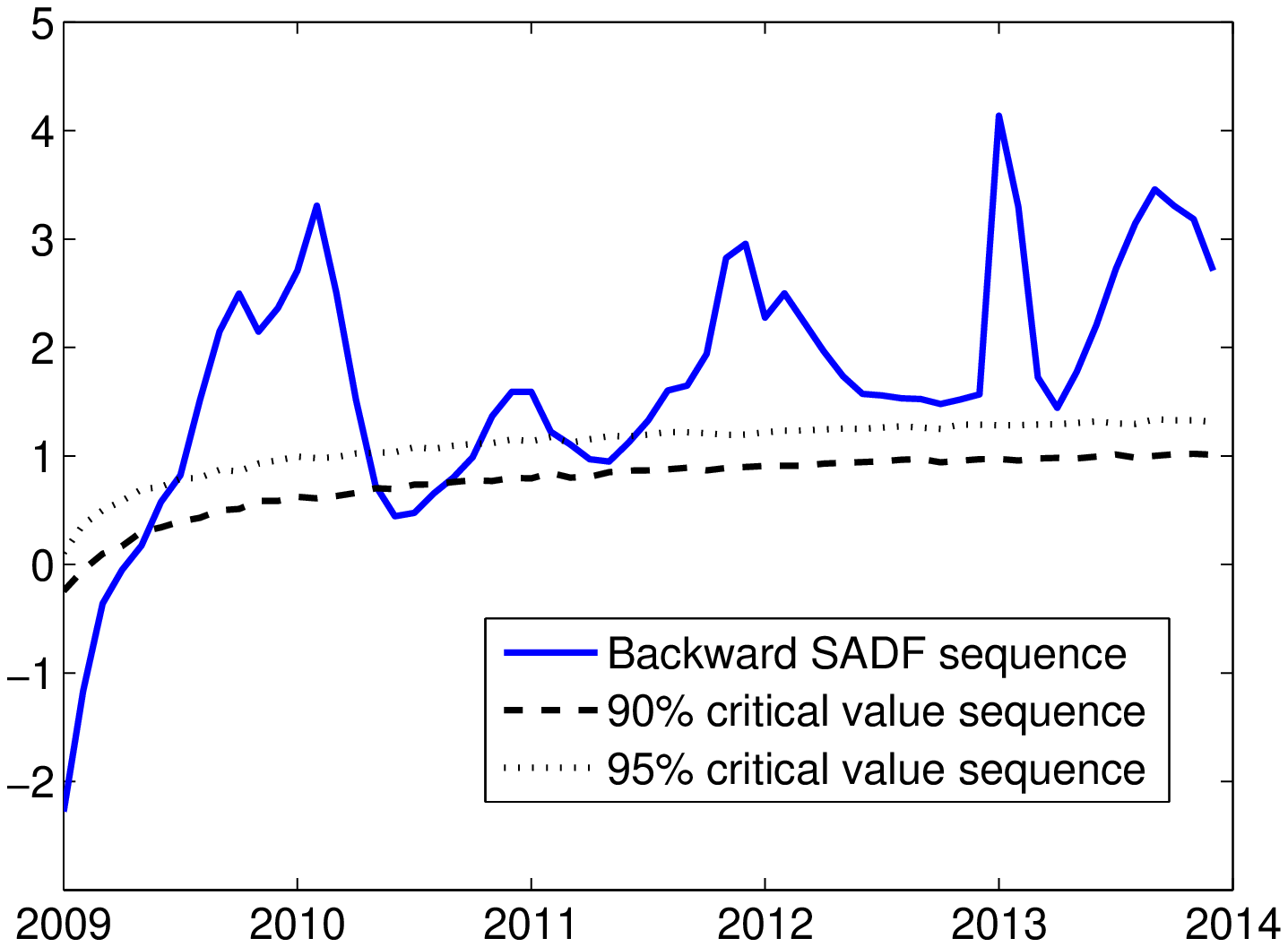}
\end{center}
\caption{Recursive calculation of the GSADF test for S\~ao Paulo. The doted lines represents the  95\% and the dashed lines the 90\% critical value sequences. (Color
  online).}
	\label{fig:SPsadf}
\end{figure}

Figures  \ref{fig:RJadf} and  \ref{fig:RJsadf} show the results of the recursive SADF and GSADF tests for Rio de Janeiro, respectively. The associated critical values are shown 
as doted 
{($95\%$ level)}
and dashed {($90\%$ level)}
curves. The SADF test shows a statistically significant explosive period  from mid-2010 to the end of 2012. The GSADF test identify a bubble beginning in mid-2010 from mid-2012. 

Figures \ref{fig:SPadf}  and \ref{fig:SPsadf}  show the results of the recursive SADF and GSADF tests for S\~ao Paulo, respectively. As before, 
doted 
{($95\%$ level)}
and dashed {($90\%$ level)}
curves represent the associated critical values. In this case, the SADF test shows a unique and long explosive behavior starting in mid-2009. On other hand the GSADF test is successful in identifying multiple bubble periods, with the last one beginning in mid-2011. 

The overall picture of Brazilian house price valuation provided by Figs. \ref{fig:RJadf},\ref{fig:RJsadf}, \ref{fig:SPadf} and \ref{fig:SPsadf} corroborate the existence of rational speculative bubbles in its two major cities. It is also noticeable that this confirms the preliminary results from glancing at Figure  \ref{fig:PRratio}. On the other hand, it is needed to emphasize that price-to-rent indices have obvious disadvantages and shortcomings. Although the indices provide information about the dynamics of the price-to-rent ratio over time, it can not provide information about the actual level of the price-to-rent ratio. Analysis of other indicators such as a household income index and a land price index \cite{pelaez}, that would be useful to check the robustness of our
results, were not performed due to the lack of monthly data on these indicators. 

\begin{figure}[ht]
\begin{center}
\includegraphics[clip,angle=0,scale=0.6]{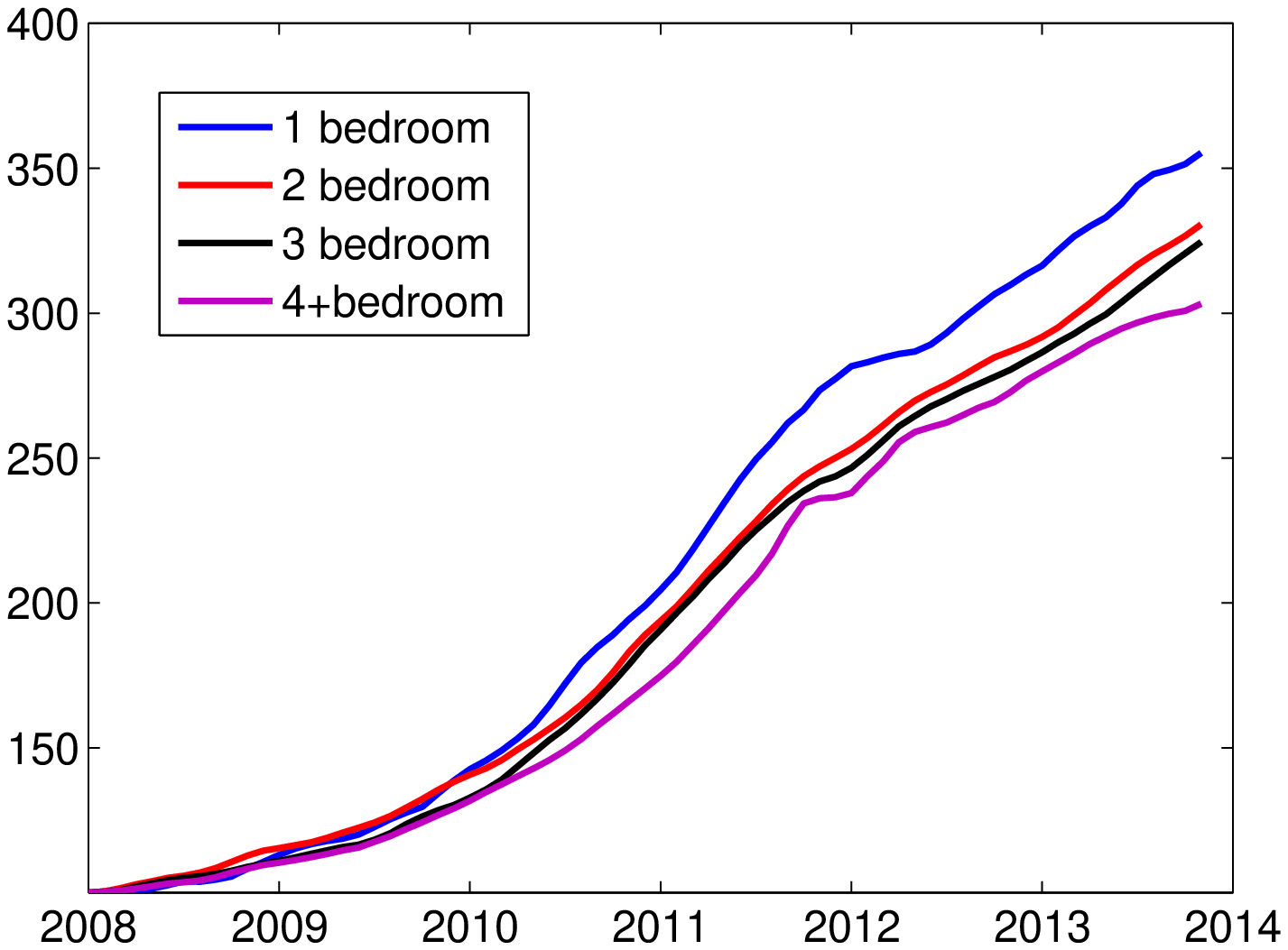}
\end{center}
\caption{Segmented price indices for Rio de Janeiro. (Color
  online).}
	\label{fig:RJq}
\end{figure}

\begin{figure}[ht]
\begin{center}
\includegraphics[clip,angle=0,scale=0.6]{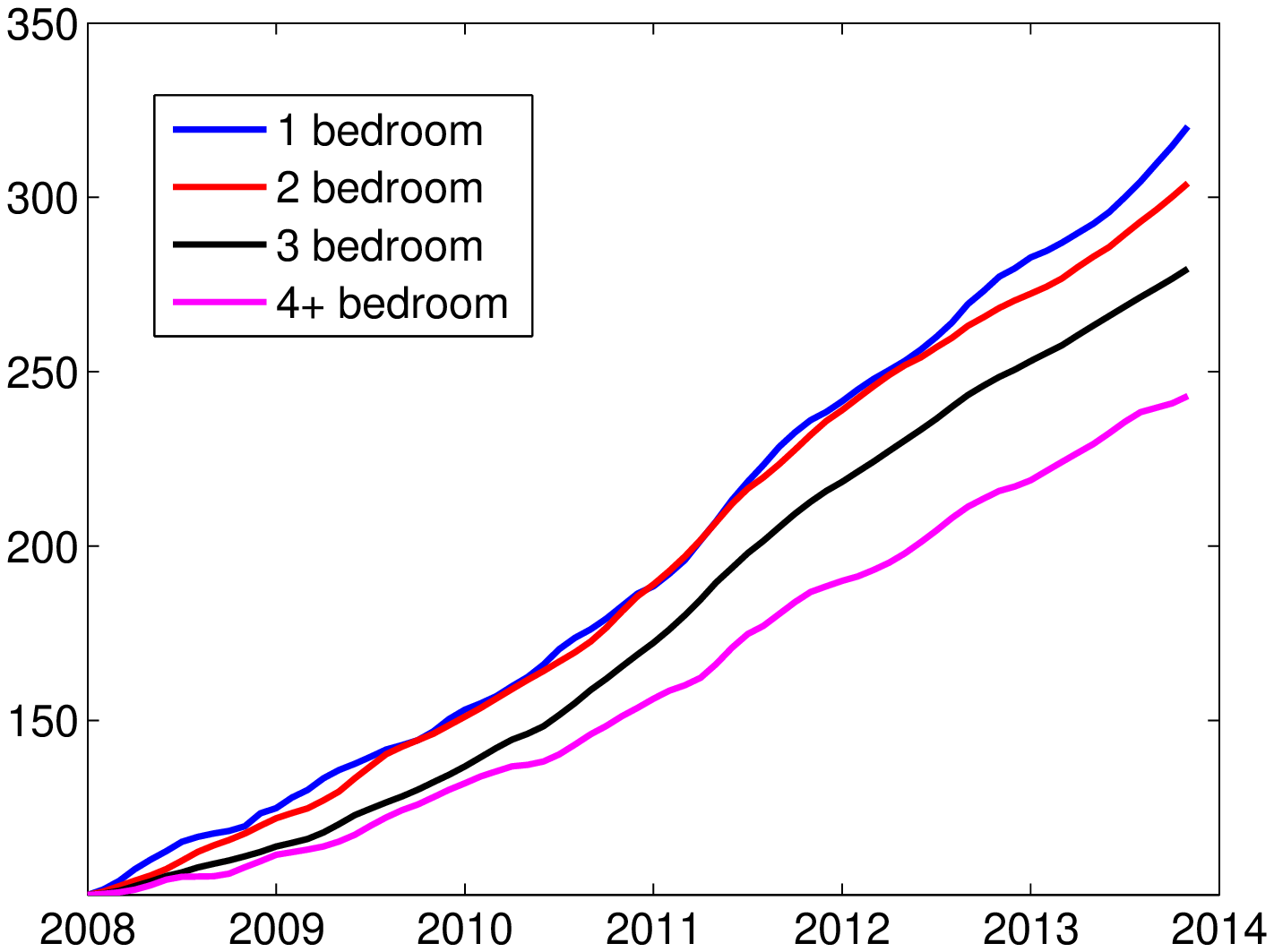}
\end{center}
\caption{Segmented price indices for S\~ao Paulo. (Color
  online).}
	\label{fig:SPq}
\end{figure}

More insight from the explosive behavior can be achieved from Figs. \ref{fig:RJq},\ref{fig:SPq}, that show the price indices segmented by the number of apartments bedrooms. Recently, based on works of  Landvoigt {\it et al.} \cite{sd} and Hott \cite{hott}, Escobari {\it et al.} \cite{escobari}
exploit the idea that low tier home prices increase at a faster pace during the
boom than the high tier home prices if cheap credit is
available to consumers predominantly at the low end of the distribution of
houses \cite{gupta} , as is the case in Brazil.

From Figure \ref{fig:RJq}, we see that until 2010 the segmented price indices grew roughly at the same pace. From 2010 to 2012, the low prices (1-bedroom) apartments index increases at a faster rate than the higher price ones. Figure \ref{fig:SPq} reveals that in S\~ao Paulo the low price apartments price increases at a considerable faster rate than high price ones, during almost all the period under consideration. These results highlight
 two findings: (i) In S\~ao Paulo, the credit for the cheap apartments consumers is an important element leading to the explosive behavior of real estate prices. On the other hand, in Rio de Janeiro, other factors may be related, as an overvaluate anticipation due to the upcoming 2014 FIFA World Cup and 2016 Olympics. (ii) In agreement with our previous findings of the PR ratio, the explosive behavior occurred in a smaller period of time in Rio than in S\~ao Paulo, which suggests that prices in Rio de Janeiro have already reached its peak. It is important to notice that this approach does not require information on market fundamentals.

\section{Conclusions}

In this paper we analyzed recent data from the Brazilian real-estate market by means of a recently proposed recursive unit-root test,  aimed at identifying explosive bubbles in real time. The test is able to identify growing bubbles and can have an important  impact on the construction of early warning systems. House prices rose dramatically in Brazil in the last few years, and our results in fact reveal the existence of speculative bubbles in the residential real estate market for the two main Brazilian cities, S\~ao Paulo and Rio de Janeiro during the recent years.


%

\begin{acknowledgements}
This work was partially supported by FAPEMIG, Brazil.
\end{acknowledgements}



\end{document}